# Guidelines for reporting cell types: the MIRACL standard

Tiago Lubiana*[1,2], Paola Roncaglia[3], Christopher J. Mungall[4], Ellen M. Quardokus[5], Joshua D. Fortriede[6], David Osumi-Sutherland[3] and Alexander D. Diehl [7]

[1] *School of Pharmaceutical Sciences, University of São Paulo, São Paulo, SP, Brazil*
[2] *Ronin Institute for Independent Scholarship*
[3] *European Bioinformatics Institute, European Molecular Biology Laboratory (EMBL-EBI), Wellcome Genome Campus, Hinxton, Cambridge, CB10 1SD, United Kingdom*
[4] *Division of Environmental Genomics and Systems Biology, Lawrence Berkeley National Laboratory, Berkeley, CA, 94720 USA*
[5] *Department of Intelligent Systems Engineering, Luddy School of Informatics, Computing, and Engineering, Indiana University, Bloomington, Indiana, USA*
[6] *Division of Biomedical Informatics, Cincinnati Children's Hospital Medical Center, Cincinnati, OH, 45224 USA*
[7] *Department of Biomedical Informatics, University at Buffalo Jacobs School of Medicine and Biomedical Sciences, Buffalo, NY, 14203 USA*

**Abstract**

Cell types are at the root of modern biology, and describing them is a core task of the Human Cell Atlas project. Surprisingly, there are no standards for reporting new cell types, leading to a gap between classes mentioned in biomedical literature and the Cell Ontology, the primary registry of cell types. Here we introduce the **M**inimal **I**nformation **R**eporting **A**bout a **C**el**L** (MIRACL) standard, a guideline for describing cell types alongside scientific articles. In a MIRACL sheet, authors organize a label, a diagnostic description, a taxon, an anatomical structure, and a parent cell class for each cell type of interest. The MIRACL standard bridges the gap between wet-lab researchers and ontologists, facilitating the integration of biomedical knowledge into ontologies and artificial intelligence systems.

## Introduction

Cell-type-oriented research has gained traction thanks to the recent blooming of projects like the Human Cell Atlas (HCA) (Rozenblatt-Rosen et al., 2017), the Human Protein Atlas (HPA) (Karlsson et al., 2021), the Human BioMolecular Atlas Program (HuBMAP) (Consortium and HuBMAP Consortium, 2019) and more (Ando et al., 2020). These efforts encompass diverse domains, from morphology to biophysics to molecular biology, and data integration requires advanced knowledge management. Programs like the Cell Annotation Platform (Osumi-Sutherland et al., 2021) and the CCF ASCT+B Reporter (Börner et al., 2021) are contributing to and reusing data annotation from the Cell Ontology (Diehl et al., 2016) to standardize references to cell types.

Despite the importance of curation, the community has yet to agree on common standards to report cell types in the scientific literature. Minimum information standards are widely used for assays and data reporting, and submission of standardized data to public databases is a prerequisite for publication for e.g. DNA and protein sequences and protein structures. Although there are standards for single-cell RNA-seq data (Füllgrabe et al., 2020) there is a broader urge to standardize metadata of cell

---

[1] Correspondence: tiago.lubiana.alves@usp.br.

type discoveries, consequently improving the organization of knowledge about cell types and states. (Aevermann et al., 2018; Bakken et al., 2017; Meehan et al., 2011; Miller et al., 2020; Panina et al., 2020)

In this perspective, we discuss the creation of guidelines for reporting cell classes, namely a Minimal Information Reporting About a CelL (MIRACL) standard focusing on the curation provided by the Cell Ontology. We outline a set of core fields for cataloging new types, and provide a schema and template for orienting reports of new cell classes.

## Results

### Who can report a MIRACL sheet?

Cell types are currently described in various research contexts, few of which are solely descriptive, microanatomy works. Any research involving a cell class not cataloged in the Cell Ontology would benefit from providing a MIRACL sheet. We outline three different situations in which a MIRACL sheet would be most valuable:

- Claims of new cell types/subtypes (Villani et al., 2017) (Joseph et al., 2021; Stecco et al., 2018). Sometimes a close superclass is cataloged, but there is value in reporting specific subclasses, as researchers might want to represent different degrees of granularity.
- New information that might change the classification of a cell type, for example the identification of a cell type in a different region. (Elmentaite et al.)
- Mentions of cell types identified in other works that are not cataloged in the Cell Ontology. (Bigaeva et al., 2020) Scenarios may include proposing a synonym for a previously reported cell type. (Popescu and Faussone-Pellegrini, 2010)

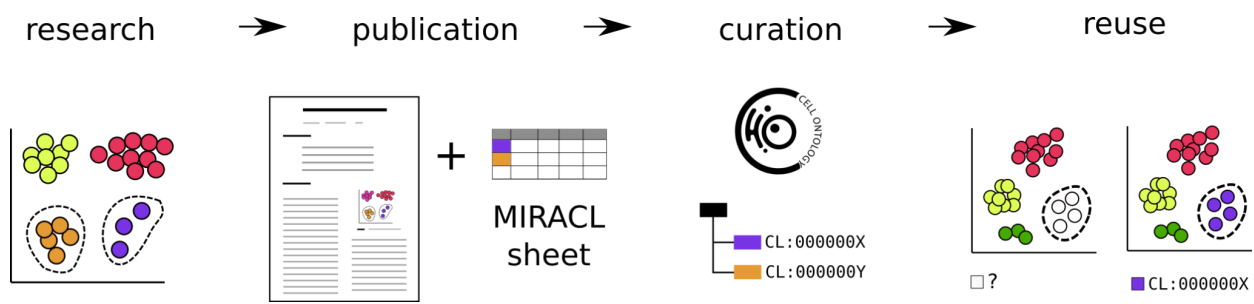

**Figure 1**: The flow from cell type discovery to knowledge reuse, with an author-guided structuring of information catalyzing the biocuration process. Currently, the curation process requires Cell Ontology editors to sift through full texts to identify the core information needed to catalog cells. A MIRACL sheet would structure information at the publication level, thereby leveraging the expertise of the original authors. That, in turn, facilitates the process of curation and, subsequently, the reuse of new terms, for example, to label clusters in single-cell omics data.

While anyone is welcome to request a new term in the Cell Ontology GitHub Tracker using standardized term templates (https://github.com/obophenotype/cell-ontology/issues), ideally, we would like to engage the authors in the process, to minimize misinterpretation of data, and reduce overhead and time required for inclusion in the Cell Ontology. If authors share information in a standardized fashion, the knowledge is quickly incorporated into the system (depicted in Figure 1), magnifying the impact of articles.

## What might a MIRACL sheet contain?

Cells are notoriously difficult to classify, as there is not a single biological feature that guides taxonomists. From simple colorimetric stains, to multiplexed immunohistochemistry, to patch clamps, to flow cytometry, to the multiple omics now performed at the single-cell level, scientists employ fundamentally different techniques to guide their perspectives on the cellular world. Any universal standard for cell type reporting faces the challenge of finding commonalities across different domains of research, while preserving the power to represent the details of each project.

A MIRACL sheet is aimed at capturing the very basic information for descriptions of new cell types. As a middle stage between full ontological axiomatization and free text, the selected fields must facilitate the work of ontology editors without requiring too much technical training of contributors. We reason that 7 different fields of information should be available for any new description of a cell type:

- A label. An unambiguous name for the cell type being described. While listing synonyms is useful, a pragmatic decision on a label is necessary when building an ontology.
- A diagnostic description. A concise, free-text description of the core characteristics of the class, similar to the diagnosis section in species taxonomy. (Winston, 1999) It should suffice to distinguish the cell type from similarly existing types (e.g. the specific marker genes).
- A superclass. A Cell Ontology class, with name and identifier, corresponding to a broader category of the new cell type. For example, a report for a new type of neuron might include "neuron (CL:0000540)" as a superclass.
- A taxon. The scientific name for the taxon for which cells of the type are expected to be found. In most cases, it will be the name of the species in which the cells were found, but this depends on the generality claim being made.
- An anatomical structure. The UBERON (Mungall et al., 2012)ontology identifier for the anatomical structure(s) in which the cells of the type were found, usually a particular organ or tissue. If cells of the type are known to span multiple structures (e.g. a long neuron), or are believed to be present in other anatomical locations, details should be noted in the additional information field. Note that some cell types, especially immune ones, may not be tied to a single anatomical structure.
- A reference. A free-text field pointing to the reference(s) that support the new cell type.
- Additional information. Any additional core information about the cell or the classification that complements the previous assertions, as well as details and technicalities that might aid curation.

In Table 1, we present an example of what a MIRACL sheet would look like in the case of two different publications (Villani et al., 2017) based on the original texts and on searching the EMBL-EBI Ontology Lookup Service for CL and UBERON terms. (Jupp et al., 2015) In a report, we propose that the information is provided in an independent tab-separated value (TSV) spreadsheet as supplementary information. Defining cell types, and attributing cell superclass(es), may sometimes be a non-trivial task. Hence, we encourage authors to provide as many suggestions as they find relevant, rather than considering the template format as a bottleneck. A ready-to-use MIRACL template in tabular format is provided in this [Google Sheet](#).

**Table 1**
An example of a MIRACL sheet

| Label | Diagnostic description | Superclass | Taxon | Anatomical structure | Reference | Additional information |
| --- | --- | --- | --- | --- | --- | --- |
| AXL$^+$ SIGLEC6$^+$ dendritic cell | A dendritic cell that expresses AXL and SIGLEC6. | dendritic cell (CL:0001056) | *Homo sapiens* | blood (UBERON:0000178) | *This article\** | - |
| Lamina propria fibroblast | A fibroblast in the lamina propria mucosa. | fibroblast (CL:0000057) | *Homo sapiens* | lamina propria (UBERON:0000030) | *This article\*, PMID: 31730855, PMID: 31474370, PMID: 30270042, PMID: 31348891, DOI:10.1101/2020.01.10.901579* | Articles used different names to refer to the class. |

\* If this sheet was provided alongside the original publications (Villani et al., 2017)(Bigaeva et al., 2020)

## MIRACL Extensions

While MIRACL's fields are designed to be simple, there are guidelines that complement MIRACL (like the Petilla convention (Petilla Interneuron Nomenclature Group et al., 2008), Allen Institute's cell type nomenclature (Miller et al., 2020), and Human Immunology Project Consortium's tentative standard (Overton et al., 2019)) that might provide additional support for particular situations. Additionally, the Cell Ontology and other OBO Foundry ontologies have tools for formal representation of other aspects of cells. To formalize, extend and make the proposal computer processable, we have started a collaborative LinkML schema (https://lubianat.github.io/miracl/).

The current may be extended by checklists applicable for particular communities, adding fields specific to subtypes of stem cell, neuron, immune cell, and others. The standard can also be tailored to particular experimental modalities, e.g. by adding a field for single-cell RNAseq markers, often used to define cell classes. (Aevermann et al., 2021)

Extensions could also adapt MIRACL to species more distant from mammals. Single-cell technology provides an opportunity to better understand plants, fungi, and bacteria. (Cole et al., 2021) MIRACL could be used with adaptations for these taxa. For example, for submission of plant cell types to the Plant Ontology (PO), the anatomical structure field would require a plant structure represented in PO, rather than an Uberon term.

## Pilot test of MIRACL sheets

To gauge community response to MIRACL standards, a pilot test was performed. The volunteers were researchers who wished to request new ontology terms for anatomical structures or cell types. Those terms were determined to be missing from either the UBERON ontology or Cell Ontology as part of their work for HuBMAP in building tables of anatomical structures, cell types and biomarkers (ASCT+B). (Börner et al., 2021)

To minimize barriers for submitting information conforming to MIRACL standards, MIRACL sheets were provided as Google Sheets templates to authors of ASCT+B tables for thymus, spleen and lymph node (v1.1, see Acknowledgements for details). Most wet bench researchers were uncomfortable engaging with the GitHub environment, where requests for anatomy and cell type terms are usually made, and the MIRACL sheets provided a simple way to obtain and communicate information directly between the experts and ontology curators. The MIRACL tables curated by ASCT+B authors are provided in the [Supplementary Data](#) as examples.

Beyond the basic MIRACL information, HuBMAP ASCT+B authors have added 2 additional fields: "Gene Biomarkers" and "Protein Biomarkers". Such extensions are welcome, as long as the core MIRACL fields are present. Additionally, the table was adapted for curating new anatomical structures for UBERON, highlighting the usefulness of the scaffold for different ontologies.

We noticed that the superclass concept was the most difficult for researchers was, while all other fields of data in the MIRACL sheet were easily provided. Ontology editors further curated the superclasses provided by the authors, and the final ontology placement may be different. Thus, if navigating the ontology proves time-consuming, authors might provide broad MIRACL superclasses (e.g. "neuron" or "lymphocyte"), as these already serve as pointers to ontology editors.

## Discussion

This work presents a standard to report new cell types, making cell discoveries explicit and accelerating curation by the Cell Ontology. MIRACL can accelerate biomedical discovery by supporting a more comprehensive support of CL, as it is used by a number of cell type annotation systems and biomedical databases (Osumi-Sutherland et al., 2021; Wang et al., 2021). In the authors' experience, scientists unfamiliar with GitHub repositories, and/or under tighter time constraints, are often reluctant to engage with the GitHub issue tracking system, but are happy to provide information in a MIRACL template to create or modify existing terms. This standard therefore can maximize output and minimize time needed for inclusion in the Cell Ontology - a gain for the entire community.

The MIRACL standard has been approved on FAIRsharing (https://fairsharing.org/4117), the major repository for biomedical standards(Sansone et al., 2019) but it only will be truly valuable if the community adopts it. MIRACL's success also depends on being adopted by journals as a requirement for reporting new cell types, and that in turn depends on a community agreement on standard operating procedures. Additionally, if researchers submit MIRACL sheets to the Cell Ontology before article submission, the cell types would be given unique identifiers that the authors (and the community) could use promptly - similarly to how e.g. accession numbers for DNA sequences can be used even before publication. To maximize the usefulness of MIRACL to end-users, we invite the reader to contact us

([https://obophenotype.github.io/cell-ontology/contact_us/](https://obophenotype.github.io/cell-ontology/contact_us/)), share opinions and contribute toward a Minimal Information Reporting About a Cell standard.

## Acknowledgments

We'd like to thank all colleagues contributing to the Cell Ontology for valuable discussions leading to this article. We also want to thank the ASCT+B authors that tested the MIRACL sheet prototype and provided feedback on the usability (Brusko et al., 2021)(Jorgensen et al., 2021a)(Jorgensen et al., 2021b) and Prof. Richard Scheuermann for insightful feedback on the preprint.

## Author contributions

Conceptualization, T.L., P.R., C.J.M., E.M.Q and A.D.D., Writing - Original Draft, T.L., Writing - Review and Editing T.L., P.R., C.J.M, E.M.Q, J.D.F., D.O.S., and A.D.D. .

## Competing interests

The authors declare no competing interests.

## Funding

This work has been made possible in part by a grant from CZI (Chan Zuckerberg Initiative DAF, an advised fund of Silicon Valley Community Foundation). TL is supported by a grant from the São Paulo Research Foundation (#19/26284-1). PR and DOS were supported by the Chan-Zuckerberg Initiative award for the Human Cell Atlas Data Coordination Platform to EMBL-EBI. JF is supported by an NHLBI/NIH award for the LungMap Data Coordinating Center (U24HL148865).

# Figure legends

**Figure 1**: The flow from cell type discovery to knowledge reuse, with an author-guided structuring of information catalyzing the biocuration process. Currently, the curation process requires Cell Ontology editors to sift through full texts to identify the core information needed to catalog cells. A MIRACL sheet would structure information at the publication level, thereby leveraging the expertise of the original authors. That, in turn, facilitates the process of curation and, subsequently, the reuse of new terms, for example, to label clusters in single-cell omics data.

# Tables

**Table 1**
An example of a MIRACL sheet

| Label | Diagnostic description | Superclass | Taxon | Anatomical structure | Reference | Additional information |
|---|---|---|---|---|---|---|
| AXL$^+$ SIGLEC6$^+$ dendritic cell | A dendritic cell that expresses AXL and SIGLEC6. | dendritic cell (CL:0001056) | *Homo sapiens* | blood (UBERON:0000178) | *This article\** | - |
| Lamina propria fibroblast | A fibroblast in the lamina propria mucosa. | fibroblast (CL:0000057) | *Homo sapiens* | lamina propria (UBERON:0000030) | *This article\*, PMID: 31730855, PMID: 31474370, PMID: 30270042, PMID: 31348891, DOI:10.1101/2020.01.10.901579* | Articles used different names to refer to the class. |

\* If this sheet was provided alongside the original publications (Villani et al., 2017)(Bigaeva et al., 2020)